\documentclass[10pt,aps,prl,twocolumn]{revtex4-1}

\usepackage{amssymb,amsmath}
\usepackage{graphicx}
\usepackage{color}
\usepackage{hyperref}
\usepackage{listings}
\usepackage{siunitx}
\usepackage{gensymb}
\usepackage{textcomp}
\usepackage{CJKutf8}

\usepackage{xspace}
\usepackage[normalem]{ulem}

\hypersetup{
    unicode=false,          
    pdftoolbar=true,        
    pdfmenubar=true,        
    pdffitwindow=false,     
    pdfstartview={FitH},    
    pdfauthor={none},     
    colorlinks=true,       
    linkcolor=blue,          
    breaklinks=true,
    citecolor=red,        
}
\usepackage{cleveref}
\graphicspath{{figures/}}

\newcommand{\ket}[1] {| #1 \rangle}

\newcommand{\micron}{{\textmu}m\xspace}
\newcommand{\Yb}{\ensuremath{^{171}{\rm Yb}^+}\xspace}
\newcommand{\twopi}{\ensuremath{2\pi \times}}

\begin{document}

\definecolor{dkgreen}{rgb}{0,0.6,0}
\definecolor{gray}{rgb}{0.5,0.5,0.5}
\definecolor{mauve}{rgb}{0.58,0,0.82}

\lstset{frame=tb,
  	language=Matlab,
  	aboveskip=3mm,
  	belowskip=3mm,
  	showstringspaces=false,
  	columns=flexible,
  	basicstyle={\small\ttfamily},
  	numbers=none,
  	numberstyle=\tiny\color{gray},
 	keywordstyle=\color{blue},
	commentstyle=\color{dkgreen},
  	stringstyle=\color{mauve},
  	breaklines=true,
  	breakatwhitespace=true
  	tabsize=3
}

\title{Quantum Gates on Individually-Addressed Atomic Qubits \\ Subject to Noisy Transverse Motion}

\author{M. Cetina}
\email{mcetina@umd.edu}
\affiliation{Joint Quantum Institute, Center for Quantum Information and Computer Science, and Departments of Physics and Electrical and Computer Engineering, University of Maryland, College Park, MD 20742}
\author{L. N. Egan}
\affiliation{Joint Quantum Institute, Center for Quantum Information and Computer Science, and Departments of Physics and Electrical and Computer Engineering, University of Maryland, College Park, MD 20742}
\author{C. A. Noel}
\affiliation{Joint Quantum Institute, Center for Quantum Information and Computer Science, and Departments of Physics and Electrical and Computer Engineering, University of Maryland, College Park, MD 20742}
\author{M. L. Goldman}
\altaffiliation[]{Now at IonQ, Inc.}
\affiliation{Joint Quantum Institute, Center for Quantum Information and Computer Science, and Departments of Physics and Electrical and Computer Engineering, University of Maryland, College Park, MD 20742}
\author{A. R. Risinger}
\affiliation{Joint Quantum Institute, Center for Quantum Information and Computer Science, and Departments of Physics and Electrical and Computer Engineering, University of Maryland, College Park, MD 20742}
\author{\begin{CJK*}{UTF8}{gbsn}D. Zhu(朱岱巍)\end{CJK*}}
\affiliation{Joint Quantum Institute, Center for Quantum Information and Computer Science, and Departments of Physics and Electrical and Computer Engineering, University of Maryland, College Park, MD 20742}
\author{D. Biswas}
\affiliation{Joint Quantum Institute, Center for Quantum Information and Computer Science, and Departments of Physics and Electrical and Computer Engineering, University of Maryland, College Park, MD 20742}
\author{C. Monroe}
\affiliation{Joint Quantum Institute, Center for Quantum Information and Computer Science, and Departments of Physics and Electrical and Computer Engineering, University of Maryland, College Park, MD 20742}
\date{\today}

\begin{abstract}

Individual trapped atomic qubits represent one of the most promising technologies to scale quantum computers, owing to their negligible idle errors and the ability to implement a full set of reconfigurable gate operations via focused optical fields.
However, the fidelity of quantum gate operations can be limited by weak confinement of the atoms transverse to the laser. 
We present measurements of this effect by performing individually-addressed entangling gates in chains of up to 25 trapped atomic ions that are weakly confined along the chain axis. We present a model that accurately describes the observed decoherence from the residual heating of the ions caused by noisy electric fields. We propose to suppress these effects through the use of ancilla ions interspersed in the chain to sympathetically cool the qubit ions throughout a quantum circuit.

\end{abstract}

\maketitle

The central challenge to scaling a quantum computer is maintaining high-fidelity coherent quantum operations while growing the number of qubits in the system. 
Isolated atomic qubits can exhibit negligible idle decoherence and near-perfect replication, so their scaling may be limited only by the external classical control. 
For example, individual atomic ions  can be confined using electromagnetic fields and their qubits can be universally controlled with light \cite{wineland2008entangled,monroe2013scaling}.  Entangling quantum gates have been demonstrated between isolated pairs of trapped ions with fidelities exceeding $99.9\%$ \cite{Ballance:2016, Gaebler2016high}.
In this work and Ref. \cite{Chapman2019pqc}, ion trap quantum computing systems have been extended to more than 20 qubits by trapping linear chains of ions.
Scaling these systems to hundreds of qubits will likely require a modular architecture featuring the shuttling of ions through a multizone ion trap chip \cite{Kielpinski:2002, pino2020demonstration} or connecting modules through photonic interconnects \cite{Monroe:2014}. 

In trapped ion systems, entanglement between qubits in a chain is typically generated via a qubit-state-dependent optical force that drives normal modes of collective oscillations \cite{molmer_multiparticle_1999, milburn2000ion, solano1999deterministic}. 
For large chains of trapped ion qubits, this requires the optical addressing of individual qubits. Here, control of the collective motion is possible via laser pulse-shaping \cite{zhu2006trapped, DebnathQC:2016}, allowing fully-connected and reconfigurable quantum gate operations \cite{Landsman2019arb, Wright:2019}.

In this Letter, we identify a limiting source of
control noise relating to the individual addressing of large chains of atomic ion qubits.
We present measurements and a model of the induced decoherence. We note that the observed effect may also limit gate fidelities in arrays of individually-addressed neutral atoms \cite{Saffman:2019}.
Finally, we suggest that even this noise source can be suppressed by sympathetically cooling the qubits, and we outline a proposal to do so by co-trapping multiple Yb isotopes \cite{Larson1986sympathetic,Chou2010frequency}.

The fidelity of entangling quantum gates between trapped ion qubits relies on the control of normal modes of motion. It is therefore common to use high-frequency radial modes to mediate the entanglement \cite{zhu2006trapped}, since they are more easily laser-cooled to the ground state and are less susceptible to heating from electric field noise \cite{Turchette:2000,Brownnutt2015}. However, heating of spectator modes can also degrade quantum gate fidelity \cite{lin2009large}.  When addressing individual ions in long chains, the motion of the ions in the weakly-confined axial direction in particular can spoil the coupling of the ions to tightly-focused individual-addressing laser beams.

We consider the effect of axial ($\hat{x}$) motion of a chain of trapped ions, each of mass $M$, with axial normal mode frequencies $\omega_m$. The ions are addressed by an array of focused laser beams that drive Rabi oscillations between two qubit states ($|0\rangle$ and $|1\rangle$). The beams are directed perpendicular to $\hat{x}$ so that, throughout their axial motion, the ions experience fixed phases of the Rabi drive.  The instantaneous qubit Rabi frequency $\Omega_i$ of the ion $i$ is proportional to the electric field amplitude of the laser beam at the position of this ion \cite{wineland2008entangled,DebnathQC:2016}.
Taking the spatial spread of each ion to be much smaller than the size of its laser beam and $\Omega_i\ll\omega_m$, the ion $i$ will experience a time-averaged Rabi frequency
\begin{equation}
\overline{\Omega}_i=\Omega_{i,0}+\frac{1}{2}\Omega''_{i,0}\sum_{m=0}^{N-1} b_{im}^2 \frac{E_m}{M\omega_m^2}, \label{eqn:avgRabi}
\end{equation}
where $\Omega_{i,0}$ is the Rabi frequency at the equilibrium position of ion $i$ and $\Omega''_{i,0}$ is its curvature along the $\hat{x}$-axis. The sum in Eq. \ref{eqn:avgRabi} is the mean-squared displacement of the $i$-th ion from its equilibrium, with  $b_{im}$ the  participation of this ion in axial mode $m$ ($\sum_m b_{im}b_{jm} = \delta_{ij}$) \cite{james1998quantum}. Here,  the energy $E_m$ in mode $m$ is assumed to be constant during a Rabi oscillation or single gate operation. 

Since cooling and heating are incoherent processes, we assume that the energies $E_m$ follow a thermal Boltzmann distribution at temperatures $T_m$, and we write $k_B T_m = \hbar \omega_m \bar{n}_m$ with $\bar{n}_m \gg 1$ the average axial vibrational occupancy number of mode $m$. If the qubit $i$ starts in $\left|0\right\rangle$ and is driven resonantly, the probability to find this qubit in $|1\rangle$ is then
\begin{equation}
\label{eqn:peModel}
p_{\ket{1}}\left(t\right)=\left\langle \sin^2 \left(\frac{\overline{\Omega}_i t}{2}\right)\right\rangle =
\frac{1 - C \cos\left(\Omega_{i,0}t+\phi\right)}{2},
\end{equation}
where $\langle\cdot\rangle$ denotes the thermal average over the energies $E_m$. The Rabi oscillations exhibit a phase advance of $\phi=\sum_m\arctan\left(\Omega_{i,0}\theta_{im} t\right)$ and a loss in contrast by a factor of $C = \prod_{m}(1+\theta_{im}^2\Omega_{i,0}^2 t^2)^{-1/2}$, where the decay parameter for ion $i$ due to mode $m$ is defined as 
\begin{equation} \label{eqn:theta}
\theta_{im}
=-\frac{k_{B}T_{m}b_{im}^{2}}{2M \omega_{m}^{2}} \frac{\Omega_{i,0}''}{\Omega_{i,0}}
 =  - b_{im}^{2}\xi_m^2 \frac{\Omega''_{i,0}}{\Omega_{i,0}}\bar{n}_m.
\end{equation}
Here, $\bar{n}_m$ is the average vibrational occupancy number of mode $m$ and $\xi_m=\sqrt{\hbar/(2 M \omega_m)}$ is the zero-point spatial spread of the ion motion (for a single trapped \Yb ion at $\omega_0 = 2\pi \times 100$~kHz, $\xi_0=17$~nm).

\begin{figure}[t!]
\centering
\includegraphics[]{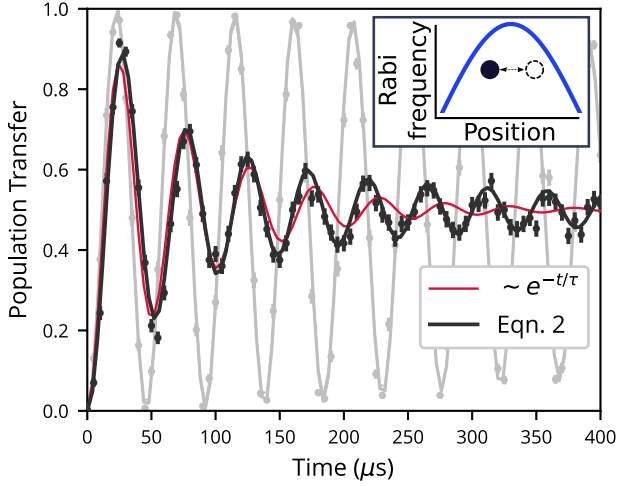}
\caption{Rabi oscillations between the ground hyperfine clock states of a single trapped \Yb ion with axial frequency 140 kHz (black) and 710 kHz (gray). The ion is prepared in the state $\ket{0}$ and driven by a pair of 355-nm Raman beams. The black and gray lines correspond to fits to Eq. \ref{eqn:peModel}. The red line corresponds to a fit to a two-state model with phase damping. Inset: illustration of the axial motion of the ion in the tightly-focused Raman beam.\label{fig:fig1}}
\end{figure}

To probe the decoherence caused by the thermal axial motion of the qubit, we confine a single \Yb ion in a microfabricated surface-electrode linear rf ion trap (HOA-2.1.1 by the Sandia National Laboratories \cite{Maunz2016}), obtaining radial secular frequencies of about 3 MHz. We address this ion with a pair of pulsed 355-nm laser beams that drive the $\rm{F}=0$  ($|0\rangle$) $\rightarrow$ $\rm{F}=1, m_F=0$ ($|1\rangle$) ground-state hyperfine  transition via a Raman process. The Raman beams are perpendicular both to each other and to the trap axis $\hat{x}$. One of the two beams is tightly focused while the other is more than ten times larger. 
Following Raman sideband cooling \cite{leibfried2003quantum} of the radial motion of the ion to an average occupation number $\bar{n}$ of less than 0.15 quanta, we optically pump the ion into $|0\rangle$. After a 5-ms delay, we use the Raman beams to drive carrier Rabi oscillations of the ion, obtaining the data shown in Fig. \ref{fig:fig1}. We observe that, when the axial frequency of the ion is decreased from 700 kHz to 140 kHz, the Rabi oscillations exhibit a sharp decay. The observed oscillations do not exhibit an exponential decay with a constant phase shift, as would be expected from pure phase damping \cite{MikeAndIke}, but instead agree with the model from Eqs. \ref{eqn:peModel}-\ref{eqn:theta}.

\begin{figure}[t!]
\centering
\includegraphics[]{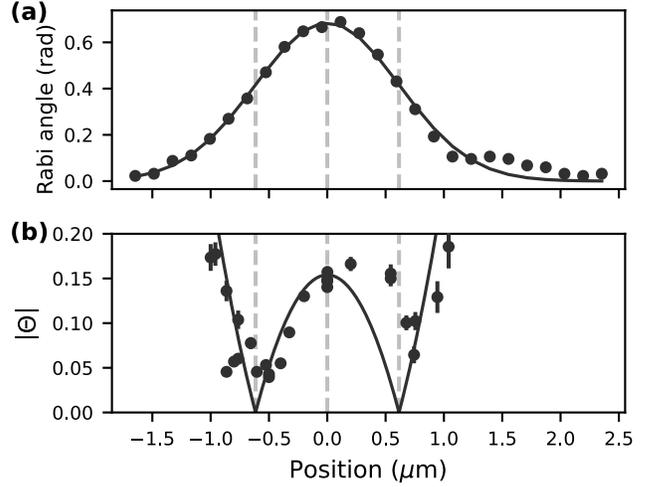}
\caption{(a)~The angle of the Rabi oscillation of a single \Yb ion driven by a 5-$\mu$s Raman pulse depending on the position of the ion in a trap with tight axial confinement ($\omega_0=\twopi$630 kHz). (b)~The absolute value of the decay parameter of Rabi oscillations of a weakly axially confined ion ($\omega_0=\twopi$140 kHz) depending on the position of the ion, together with a prediction based on the Gaussian fit from Fig. 2(a) and an axial ion temperature corresponding to $\bar{n} = 280$ quanta. 
\label{fig:fig2}}
\end{figure}

We investigate the spatial dependence of the decoherence by applying static trap voltages to move the ion along $\hat{x}$. First, we map out the spatial profile of our tightly-focused Raman beam by tightly confining the ion and then driving it with a carrier Raman pulse of duration $\tau=0.5$~$\mu$s. We use the fraction of the ion population transferred to $|1\rangle $ to determine the Rabi angle $\Omega_{1,0} \tau$, as shown in \crefformat{figure}{Fig.~#2#1{(a)}#3}\cref{fig:fig2}. 
A Gaussian fit to the obtained data ($\Omega_{1,0}\tau\sim e^{-x^{2}/w^{2}}$) yields the 1/$e^2$ intensity-radius of the tightly focused beam $w=$ 870(25)~nm. On one shoulder of the tightly focused beam, we observe a deviation from the Gaussian shape, which we ascribe to the mode profile of the laser. 

Next, we relax the single ion's axial confinement and perform a carrier Rabi oscillation experiment, as in Fig. \ref{fig:fig1}, at each set position of the ion. We fit the resulting data to extract the decay parameter $\theta \equiv \theta_{10}$, with the results shown in \crefformat{figure}{Fig.~#2#1{(b)}#3}\cref{fig:fig2}. We compare the obtained  $\theta(x)$ data to the prediction 
$\theta  = 2(\xi_0/w)^2\left(1-2x^{2}/w^{2}\right)\bar{n}_0$, 
which assumes a Gaussian shape of the tightly-focused beam, with the ion's axial 
average thermal vibrational number of $\bar{n}_0 = 280$ at the expected Doppler cooling
temperature. 
We observe good qualitative agreement after accounting for the laser's mode shape. In striking contrast to decoherence caused by beam fluctuations,  which increases on the sides of the focused beam, the measured decay parameter reaches its minimum values near the inflection points of the Gaussian curve. The measurements from Figs. \ref{fig:fig1} and \ref{fig:fig2} demonstrate that both the spatial and the temporal behavior of the observed Rabi oscillations of an ion under weak axial confinement are well described by the model of Eqs. \ref{eqn:peModel}-\ref{eqn:theta}.

\begin{figure}[t!]
\centering
\includegraphics[]{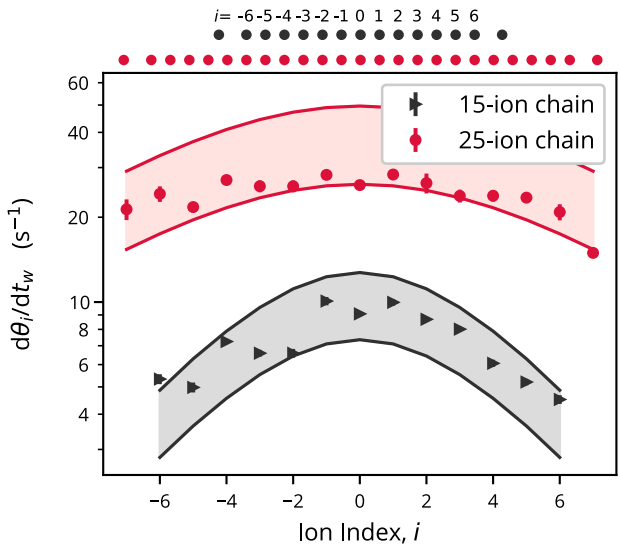}
\caption{The rate of change of the decay parameter $\theta_i$ as function of the ion index ($i$) in a chain of near-equispaced 15 ions (black) and 25 ions (red). The error bars are statistical from fits of $\theta_i(t_w)$ to a linear increase with the wait time $t_w$. The solid lines correspond to predictions based on power-law spectral density of electric field noise with exponent $\alpha=[0.8,1]$ and the independently measured heating rate of one ion at 3 MHz of $\dot{\bar{n}}_r = 88(6)$ quanta/s. 
\label{fig:fig3}}
\end{figure}

We now consider the effect of axial motion on an array of atomic ions.
After sufficient time $t_{\rm{w}}$ following laser cooling, the temperature $T_m$ of the axial mode $m$ will be dominated by the work done by noisy background electric fields \cite{Turchette:2000, Brownnutt2015}. If this field is uniform in space, its work on mode $m$ will be proportional to $\left(\sum_i b_{i,m}\right)^2$. 
Since the contribution of the $m$-th mode to the decay parameter scales as $\omega_{m}^{-2}$, and, for higher-frequency modes, $b_{i,m}$ oscillates with the ion index $i$ (i.e., neighboring ions move out of phase with each other), we expect the heating of the lowest-frequency ``in-phase" axial mode ($m=0$), to strongly dominate the gate error budget. In this case, the effects of the axial motion on the $i$-th ion are captured by the single decay parameter
\begin{equation}\label{eqn:COMheatModel}
\theta_{i} \equiv \theta_{i0} = b_{i0}^2\left(\sum_j b_{j0}\right)^2\theta(t_{\rm{w}}),
\end{equation}
where $\theta(t_{\rm{w}})$ is the decay parameter of a single ion in a trap with axial frequency equal to $\omega_0$, following Eq. \ref{eqn:theta}. Note that for harmonic axial confinement where the in-phase mode is the center-of-mass mode ($b_{i0}=N^{-1/2}$), the decay parameter  $\theta_{i}$ becomes equal to that for a single ion.

In the experiment, we use a combination of quadratic and quartic axial potentials \cite{lin2009large} to prepare near-equispaced chains of 15 (25) \Yb ions with 4.4~\micron ion spacing, obtaining 193 (123)~kHz as the lowest axial mode frequency. We use a 32-channel acousto-optic modulator \cite{DebnathQC:2016} to produce individually-controlled, tightly-focused 355-nm beams identical to the one from Fig. \ref{fig:fig2}. Following sideband cooling of the radial modes of the ion chain, we drive simultaneous Raman Rabi oscillations on the middle 13 (15) ions in the chain after a variable wait time $t_{\rm{w}}$ following the Doppler and sideband cooling. We fit the oscillations of each ion to the model from Eqs.  \ref{eqn:peModel}-\ref{eqn:theta} to determine the rate of change of its decay parameter, and we show the results of these measurements in Fig. \ref{fig:fig3}. The observed variation of the decay parameters across the chains follows the factor  $b_{i0}^2$ from Eq. \ref{eqn:COMheatModel}, with the Rabi oscillations of the middle ions exhibiting increased decay due their  higher participation in the lowest-frequency axial mode. 

Electric-field noise in ion traps is empirically observed to follow a power-law with frequency $\omega^{-\alpha}$, with exponent $\alpha$ between 0 and 2 \cite{Brownnutt2015}. To check for consistency of our observations with this behavior, we use sideband spectroscopy of a single ion to  measure the heating rate of $\dot{\bar{n}}_r=88(6)$ quanta/s for a 3-MHz radial mode parallel to the trap surface. 
In Fig. \ref{fig:fig3}, we show the predictions based on Eqs. \ref{eqn:theta} and \ref{eqn:COMheatModel} and $\alpha = 1$. We also show predictions for $\alpha=0.8$, which we deduced independently from $d\theta/dt_{\rm{w}}$ for a single ion as a function of the axial trap frequency \cite{MCSupMat}. We observe good quantitative agreement with our data, suggesting that electric-field noise in our system is consistent with previously observed values.

The Rabi frequency $\Omega_{i,j}$ of the entangling dynamics between the $|00\rangle$ and $|11\rangle$ states of ions $i$ and $j$ during two-qubit entangling gates \cite{molmer_multiparticle_1999, milburn2000ion, solano1999deterministic} is proportional to the product of the single-qubit Rabi frequencies on the two addressed ions. If the two ions are centered on the maxima of their respective individual-addressing beams, the joint decay parameter corresponding to axial mode $m$ is $\theta_{im} + \theta_{jm}$. The fidelity of the obtained two-qubit state is bounded from above by 
$F_{ij} = {\rm Tr}\left(\hat{\rho}\left(t\right)\hat{\rho}_\chi\right) = \left\langle \cos^{2}\left(\bar{\Omega}_{i,j}t-\chi\right)\right\rangle$, where $\chi$ is the desired two-qubit gate angle \cite{molmer_multiparticle_1999}, and $\bar{\Omega}_{i,j}$ is the mean value of $\Omega_{i,j}$ during the ions' axial motion. Performing the thermal average as in Eq. \ref{eqn:peModel}, we find the state fidelity bound after $N_g$ successive fully-entangling gates ($\chi=N_g\pi/4$) is \begin{equation}\label{eqn:MS_F}
F_{ij}=\frac{1}{2}+\frac{1}{2}\prod_{m}\frac{1}{\sqrt{1+\left(N_g \pi/2\right)^{2}(\theta_{im}+\theta_{jm})^{2}}}.
\end{equation}

To check this prediction, we apply amplitude modulation to the focused Raman beams addressing two ions in the above chains of 15 (25) ions so as to perform one or three successive 250 $\mu$s (500 $\mu$s)-long entangling gates between the target ions \cite{DebnathQC:2016}, after a variable wait time $t_{\rm{w}}$ following Doppler and sideband cooling.
After applying the entangling gate(s), we measure the $\{|00\rangle, |11\rangle\}$ subspace population $p_{00}+p_{11}$. Separately, we apply additional $\pi/2$ pulses with variable phase to both ions and extract the parity fringe contrast $C$, to witness entanglement \cite{sackett2000experimental}. We independently determine $\theta_i$ and $\theta_j$ in our chains as a function of $t_w$ by repeating the measurements from Fig. \ref{fig:fig3} and fitting the resulting decay parameters to a linear increase with the wait time $t_w$ \cite{MCSupMat}.

\begin{figure}[t!]
\centering
\includegraphics[]{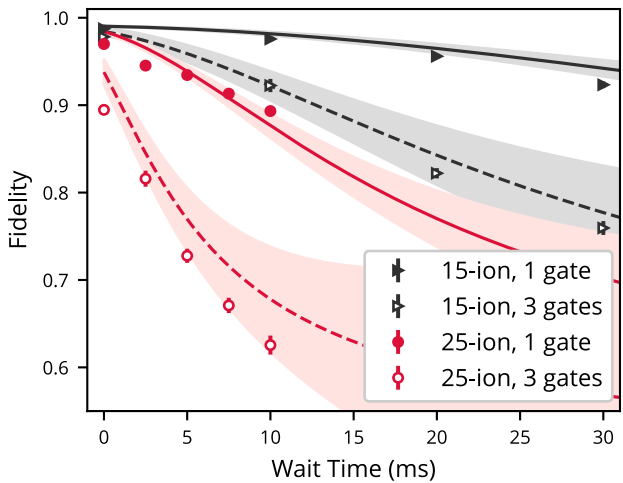}
\caption{The fidelity of one (three) fully entangling gate(s) on two ions (indices $i$=$-$6 and $j$=$-$5 as in Fig. \ref{fig:fig3}) in a chain of 15 and 25 ions as a function of the waiting time before the gate(s).  The error bars denote $1\sigma$ uncertainty of the weighted average of several measurements of the fidelity. The shaded areas show the predictions based on Eq. \ref{eqn:MS_F}, adjusted down by the error in state preparation and measurement in our ion chains (0.9$\%$, \cite{MCSupMat}). The uncertainty in the theory predictions reflect the errors in our determination of the decay parameters $\theta_i$ and $\theta_j$.
\label{fig:fig4}}
\end{figure}

We compute the gate fidelities from our measurements as $F=(p_{00}+p_{11}+C)/2$ \cite{sackett2000experimental} and compare these in Fig. \ref{fig:fig4} to the predictions based on Eq. \ref{eqn:MS_F}, with only the lowest ($m=0$) axial mode contributing. We observe good agreement between our measurements and the model of Eq. \ref{eqn:MS_F}. In particular, in chains of 15 (25) ions, after $t_w$=10 (2.5) ms, corresponding to the time it takes to complete 40 (5) sequential individual entangling gates, our model explains most of the observed loss in gate fidelity. 

In ion chains, $\omega_0$ is roughly inversely proportional to the ion number $N$ \cite{Schiffer1993pti,MCSupMat}. Assuming electric field noise with exponent $\alpha$ and using Eqs. \ref{eqn:theta}-\ref{eqn:COMheatModel}, we obtain $d\theta_i/dt_w\sim N^{2+\alpha}$ in such chains, implying that the entangling gate error scales as $t_w^2 N^{4+2\alpha}$. For $\alpha\approx 1$ as seen in many ion trap experiments \cite{Brownnutt2015}, this results in a scaling of gate errors proportional to $N^6$, which would limit the use of long chains with high-fidelity operations.

To address the challenge posed by axial heating, we propose to intersperse the coherent operations with periodic sympathetic cooling \cite{Larson1986sympathetic} of axial modes via coolant ions that are distributed throughout the ion chain \cite{Larson1986sympathetic, lin2009large, Chou2010frequency, pino2020demonstration}.  
Since the gate error is proportional to $t_w^2$, in the limit of frequent cooling to the same initial axial temperature $T_0$, we expect sympathetic cooling to strongly suppress of the effects of axial heating on gate fidelity. 

We propose to apply the \Yb narrow-line sideband cooling scheme from Ref. \cite{Revelle2019dos} to even-isotope Yb$^+$ ions that are co-trapped in a linear chain with \Yb. Here, $\pi$-polarized 435-nm light addresses the red motional sidebands of the $^2$S$_{1/2}$-$^2\rm{D}_{3/2}$ transition, while a 935-nm laser resets the ions' state via the  $^2\rm{D}_{3/2}$-$^{3}\rm{D}[3/2]_{1/2}$ transition. Using a  narrow-linewidth 435-nm laser, the dominant crosstalk to the \Yb qubits would arise from off-resonant absorption of the spontaneously-emitted  photons on the $^{3}\rm{D}[3/2]_{1/2}$-$^{2}\rm{S}_{1/2}$ transition ($\lambda$ = 297~nm). Given the near-unity branching ratio of this transition \cite{olmschenk2007manipulation}, the average 297-nm excitation rate per qubit, for an arbitrarily long \Yb chain and unity saturation of the 935-nm transition, will be at most $R=r\lambda^2(\Gamma/2)^3/(16 d^2 \Delta^2)$, with $r$ the fraction of coolant ions,  d $\approx$ 4~$\mu\rm{m}$ the ion spacing,  $\Gamma=2\pi\times$4.3 MHz the linewidth of the $^{3}$D[3/2]$_{1/2}$ state  \cite{Berends1993bll}, and $\Delta$ the isotope splitting on the 297-nm transition. Using a King plot of the even-isotope spectroscopic data from \cite{Meyer2012lsa}, together with the hyperfine splittings in \Yb from \cite{olmschenk2007manipulation}, we establish that, for the most common isotopes $^{172,174}\rm{Yb}^{+}$,  $\Delta>2\pi\times$ 2.4 GHz and, for $r=0.5$,  $R<2\times10^{-3}/\rm{s}$. Since only the first several axial modes need to be cooled,  fewer coolant ions and a lower cooling duty cycle would likely suffice. Moreover, sideband-cooling is a dark-state cooling scheme, and some of the entropy-removing spontaneous scatterings are elastic. Therefore, the true error rate is likely at least an order of magnitude smaller. These estimates suggest that our technique could mitigate the impact of axial heating without itself becoming the limiting factor in the system's performance.
 
 We acknowledge fruitful discussions with M. Revelle, J. Kim, M. Li, N. Pisenti, and K. Wright and the contributions of J. Mizrahi, K. Hudek, J. Amini and K. Beck to the experimental setup. 
 This work is supported by the ARO through the IARPA LogiQ program under 11IARPA1008, the NSF STAQ Program, the AFOSR MURIs on Quantum Measurement/Verification and Quantum Interactive Protocols, the ARO MURI on Modular Quantum Circuits, DOE BES award de-sc0019449, DOE HEP award de-sc0019380, and the NSF Physics Frontier Center at JQI. L. Egan is also funded by NSF award DMR-1747426.

\section{Supplemental Material}

\subsection{Frequency Dependence of Heating}

Here we present measurements of the rate of change of the single ion decay parameter $\theta$ as a function of the axial trap frequency $\omega_0$. We change the voltages applied to the trap to vary $\omega_0$. At each voltage setting, we record carrier Rabi oscillations of the ion after varying wait times $t_w$. We fit these data using Eqs. 2-3 to extract $d\theta/dt_w$. We show the results of these measurements in Fig. \ref{fig:figA}. We fit the obtained values of $d\theta/dt_w$ as a function of the axial frequency $\omega_0$ to a power law. To account for the effect of radial heating, we include an offset $B$ into our model by writing $d\theta/dt_w = A \omega_0^{-2-\alpha} + B$. We observe good agreement of the fit with our data. From the fit, we obtain $\alpha = 0.8(1)$, corresponding to electric field noise with power spectral density that scales as $\omega^{-0.8(1)}$. We obtain $B=0.9(1)/\rm{s}$, corresponding to $\dot{\bar{n}}_r = 150(20)/$s, which agrees with the experimentally measured radial heating for modes oriented at $45^{\degree}$ to the trap surface.

We note that the mean rate of increase in the decay parameter for 15 (25) ions is 1.2 (1.35) times larger than the prediction of  Eq. 4. We attribute this to heating of additional modes of motion (not just the in-phase $m=0$ mode), driven by the inhomogeneity of the heating electric field arising from the 70-$\mu$m distant trap surface across the 62 (106) $\mu$m-long ion chain.
\renewcommand{\thefigure}{S\arabic{figure}}
\renewcommand{\thetable}{S\Roman{table}}
\renewcommand{\theequation}{S\arabic{equation}}
\setcounter{equation}{0}
\setcounter{figure}{0}
\begin{figure}[h!]
\centering
\includegraphics[]{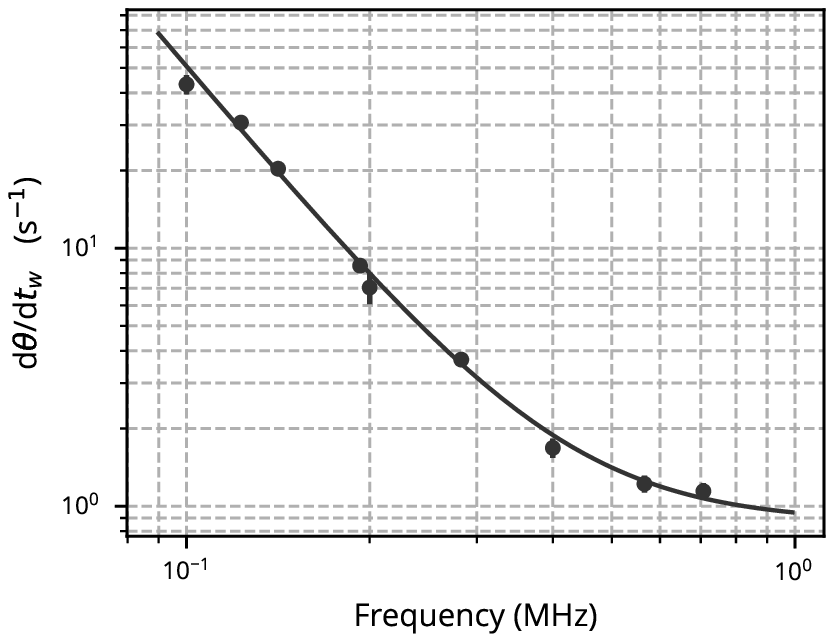}
\caption{The rate of change of the decay parameter $\theta$ as function of frequency. The error bars correspond to the 1$\sigma$ fit estimates. 
\label{fig:figA}}
\end{figure}

\subsection{State Preparation and Measurement Errors}

We characterize state preparation and measurement (SPAM) errors for 2-qubit gates in a chain of 15 ions by measuring the states of the ions $i=-6,-5$ (see Fig. 3 in the main text) following different state preparations. The qubits are prepared in the states $\ket{00}, \ket{01}, \ket{10},~\text{and}~\ket{11}$, with $2\times10^4$ trials in each state, and measured in the same basis. We initialize qubits to state $\ket{0}$ using optical pumping; to prepare a qubit in state $\ket{1}$ we rotate the initialized qubit by $\pi$ using an SK1 pulse \cite{merrill2014sk1}. To measure the qubit state, we shine 369-nm light, which is resonant with the transition between the \{$^2$S$_{1/2}, \rm{F} = 1$\} and \{$^2$P$_{1/2}, \rm{F} = 0$\} manifolds. Ions scatter photons when they are in the $\ket{1}$ (bright) state but not in the $\ket{0}$ (dark) state. Using individual PMTs to detect scattered photons from each ion for $100~\mu$s, we determine that the ion is bright (dark) when a  measurement generates $>1$ ($\leq 1$) photons. 

\begin{table}[h!]
\begin{tabular}{llcccc}
 &  &  & \multicolumn{1}{r}{\textbf{measured}} & \multicolumn{1}{l}{\textbf{state} $\vspace{.15cm}$} & \multicolumn{1}{l}{} \\
 & $\vspace{.15cm}$ & \multicolumn{1}{l}{$\hspace{.35cm} \ket{00} \hspace{.35cm}$} & \multicolumn{1}{l}{$\hspace{.45cm}\ket{01} \hspace{.25cm}$} & \multicolumn{1}{l}{$\hspace{.35cm}\ket{10} \hspace{.35cm}$} & \multicolumn{1}{l}{$\hspace{.35cm}\ket{11} \hspace{.35cm}$} \\
\multicolumn{1}{c}{} & $\hspace{.3cm}\ket{00}\hspace{.2cm}\vspace{.15cm}$ & 99.76\% & 0.17\% & 0.07\% & 0.00\% \\
\multicolumn{1}{c}{\textbf{prepared}} & $\hspace{.3cm}\ket{01}\hspace{.2cm}\vspace{.15cm}$ & 0.53\% & 99.34\% & 0.00\% & 0.13\% \\
\multicolumn{1}{c}{\textbf{state}} & $\hspace{.3cm}\ket{10}\hspace{.2cm}\vspace{.15cm}$ & 0.36\% & 0.00\% & 99.22\% & 0.42\% \\
 & $\hspace{.3cm}\ket{11}\hspace{.2cm}\vspace{.15cm}$ & 0.02\% & 0.45\% & 0.48\% & 99.05\%
\end{tabular}
\caption{State preparation and measurement populations computed on ions $i=-6$ and $j=-5$ (see Fig. 3 in the main text), where states are given by $\ket{\psi_i \psi_j}$ \label{tab:spam}}

\end{table}

SPAM errors are calculated (Table \ref{tab:spam}) and used to correct the predicted two-qubit gate fidelities in Fig. 4 of the main text. We emphasize that the measured two-qubit gate fidelities shown in Fig. 4 are not corrected for SPAM errors.

\subsection{Lowest Axial Frequency in Equispaced Ion Chains}

In the limit of many ions ($N\rightarrow\infty$), an equispaced ion chain can be approximated by a continuous charge distribution with linear charge density $e/d$, where $d$ is the ion spacing. The Coulomb potential of this charge is exactly countered by the applied trap potential
\begin{equation} \label{eqn:equispacedV}
V(x) = \frac{e^2}{4\pi\epsilon_0 d} \ln\frac{(N/2)^2}{(N/2)^2-(x/d)^2},
\end{equation}
which holds the ions in place. To model near-equispaced chains, we use $V(x)$ as the trap potential and numerically find the equilibrium ion positions $x_{i,0}$. We find that, when using the potential from Eq. \ref{eqn:equispacedV}, for all $N<250$, the deviation of the equilibrium ion  positions from that of an equal-spaced chain is at most $0.02~d$.

\begin{figure}[t!]
\centering
\includegraphics[]{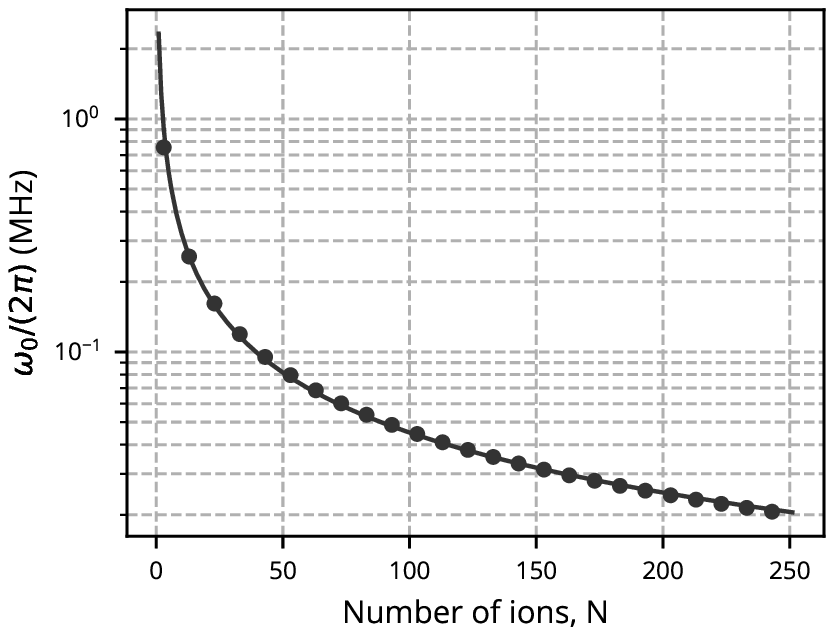}
\caption{The lowest axial frequency $\omega_0$ of a chain of $^{171}$Yb$^+$ ions as a function of the number of ions $N$ in the potential given by Eq. \ref{eqn:equispacedV} with $d=4.4$ $\mu$m. The solid line corresponds to the fitted $N^{-0.856}$ power law.
\label{fig:figB}}
\end{figure}

For ion positions $x_i$ near the equilibrium, the total energy can be written as $\frac{e^2}{2 d^3}\sum_{i,j=1}^N (x_i-x_{i,0}) Q_{i,j} (x_j - x_{j,0})$, where
\begin{equation} \label{eqn:equispacedV2}
\begin{split}
Q_{i,i} & = \frac{2(N/2)^2+2(x_{i,0}/d)^2}{((N/2)^2-(x_{i,0}/d)^2)^2} + \sum_{j\neq i} \frac{2 d^3}{|x_{i,0}-x_{j,0}|^3} \\
Q_{i\neq j} & = -\frac{2 d^3}{|x_{i,0}-x_{j,0}|^3}.
\end{split}
\end{equation}
The axial mode frequencies are then found as $\omega_m=\omega_u \sqrt{\lambda_m}$, where $\lambda_m$ are the eigenvalues of the matrix $Q_{i,j}$ and $\omega_u=\sqrt{e^2 / (4\pi\epsilon_0 M d^3)}$ is the unit frequency. The calculated frequency $\omega_0$ of the lowest-frequency axial mode is plotted as a function of the ion number, $N$, in Fig. \ref{fig:figB}. For chains of up to 250 ions, we observe good agreement with the $N^{-0.86}$ scaling that was predicted for chains in harmonic traps by Ref.  \cite{Schiffer1993pti}.

%

\end{document}